\documentclass[aps,prl,preprint,tightenlines,superscriptaddress,byrevtex,twocolumn,11pt,nofootinbib]{revtex4}
\usepackage{graphicx} 
\usepackage{dcolumn}  

\graphicspath{{ps}}

\begin{document}


\preprint{\vbox{ \hbox{   }
                 \hbox{BELLE-CONF-0807}
}}

\title{\boldmath \center Precise measurement
 of ${\cal B}(\tau^-\to K^{*0}(892) K^- \nu_\tau)$ and  \\
the mass and width of the $K^{*0}(892)$ meson} 

\affiliation{Budker Institute of Nuclear Physics, Novosibirsk}
\affiliation{Chiba University, Chiba}
\affiliation{University of Cincinnati, Cincinnati, Ohio 45221}
\affiliation{Department of Physics, Fu Jen Catholic University, Taipei}
\affiliation{Justus-Liebig-Universit\"at Gie\ss{}en, Gie\ss{}en}
\affiliation{The Graduate University for Advanced Studies, Hayama}
\affiliation{Gyeongsang National University, Chinju}
\affiliation{Hanyang University, Seoul}
\affiliation{University of Hawaii, Honolulu, Hawaii 96822}
\affiliation{High Energy Accelerator Research Organization (KEK), Tsukuba}
\affiliation{Hiroshima Institute of Technology, Hiroshima}
\affiliation{University of Illinois at Urbana-Champaign, Urbana, Illinois 61801}
\affiliation{Institute of High Energy Physics, Chinese Academy of Sciences, Beijing}
\affiliation{Institute of High Energy Physics, Vienna}
\affiliation{Institute of High Energy Physics, Protvino}
\affiliation{Institute for Theoretical and Experimental Physics, Moscow}
\affiliation{J. Stefan Institute, Ljubljana}
\affiliation{Kanagawa University, Yokohama}
\affiliation{Korea University, Seoul}
\affiliation{Kyoto University, Kyoto}
\affiliation{Kyungpook National University, Taegu}
\affiliation{\'Ecole Polytechnique F\'ed\'erale de Lausanne (EPFL), Lausanne}
\affiliation{Faculty of Mathematics and Physics, University of Ljubljana, Ljubljana}
\affiliation{University of Maribor, Maribor}
\affiliation{University of Melbourne, School of Physics, Victoria 3010}
\affiliation{Nagoya University, Nagoya}
\affiliation{Nara Women's University, Nara}
\affiliation{National Central University, Chung-li}
\affiliation{National United University, Miao Li}
\affiliation{Department of Physics, National Taiwan University, Taipei}
\affiliation{H. Niewodniczanski Institute of Nuclear Physics, Krakow}
\affiliation{Nippon Dental University, Niigata}
\affiliation{Niigata University, Niigata}
\affiliation{University of Nova Gorica, Nova Gorica}
\affiliation{Osaka City University, Osaka}
\affiliation{Osaka University, Osaka}
\affiliation{Panjab University, Chandigarh}
\affiliation{Peking University, Beijing}
\affiliation{Princeton University, Princeton, New Jersey 08544}
\affiliation{RIKEN BNL Research Center, Upton, New York 11973}
\affiliation{Saga University, Saga}
\affiliation{University of Science and Technology of China, Hefei}
\affiliation{Seoul National University, Seoul}
\affiliation{Shinshu University, Nagano}
\affiliation{Sungkyunkwan University, Suwon}
\affiliation{University of Sydney, Sydney, New South Wales}
\affiliation{Tata Institute of Fundamental Research, Mumbai}
\affiliation{Toho University, Funabashi}
\affiliation{Tohoku Gakuin University, Tagajo}
\affiliation{Tohoku University, Sendai}
\affiliation{Department of Physics, University of Tokyo, Tokyo}
\affiliation{Tokyo Institute of Technology, Tokyo}
\affiliation{Tokyo Metropolitan University, Tokyo}
\affiliation{Tokyo University of Agriculture and Technology, Tokyo}
\affiliation{Toyama National College of Maritime Technology, Toyama}
\affiliation{Virginia Polytechnic Institute and State University, Blacksburg, Virginia 24061}
\affiliation{Yonsei University, Seoul}
  \author{I.~Adachi}\affiliation{High Energy Accelerator Research Organization (KEK), Tsukuba} 
  \author{H.~Aihara}\affiliation{Department of Physics, University of Tokyo, Tokyo} 
  \author{D.~Anipko}\affiliation{Budker Institute of Nuclear Physics, Novosibirsk} 
  \author{K.~Arinstein}\affiliation{Budker Institute of Nuclear Physics, Novosibirsk} 
  \author{T.~Aso}\affiliation{Toyama National College of Maritime Technology, Toyama} 
  \author{V.~Aulchenko}\affiliation{Budker Institute of Nuclear Physics, Novosibirsk} 
  \author{T.~Aushev}\affiliation{\'Ecole Polytechnique F\'ed\'erale de Lausanne (EPFL), Lausanne}\affiliation{Institute for Theoretical and Experimental Physics, Moscow} 
  \author{T.~Aziz}\affiliation{Tata Institute of Fundamental Research, Mumbai} 
  \author{S.~Bahinipati}\affiliation{University of Cincinnati, Cincinnati, Ohio 45221} 
  \author{A.~M.~Bakich}\affiliation{University of Sydney, Sydney, New South Wales} 
% ...
\collaboration{The Belle Collaboration}

\noaffiliation

\begin{abstract}
Using the high statistics $\tau$ data {sample}
recorded {in the Belle experiment} at KEKB, 
we have greatly improved the precision of the branching fraction 
 ${\cal B}(\tau^-\to K^{*0}(892) K^-\nu_\tau) = (1.56\pm 0.02\pm 0.09)\times 10^{-3}$, 
while the mass and width of the $K^{*0}(892)$ meson are measured to be 
{$(895.10\pm 0.27\pm 0.31)$ MeV/c$^2$} and 
{$(47.23\pm 0.49\pm 0.79)$} MeV/c$^2$, respectively, 
with better accuracy than the PDG world average values. 
The first measurement of the decay $\tau^-\to K^{*0}(892) K^-\pi^0\nu_\tau$ is
 also reported with a branching fraction of 
{
${\cal B}(\tau^-\to K^{*0}(892) K^-\pi^0\nu_\tau)=(2.39\pm 0.46\pm 0.26)\times 10^{-5}$. }
\vspace{1pc}
\end{abstract}


\maketitle

\tighten
\section{Introduction}

In addition to high statistics
samples of B mesons, B factories
also provide large number of $\tau$-leptons, which can be used for
high precision measurements.
We report here a study of the decay $\tau^-\to K^{*0}(892) K^-\nu_\tau$,
{in which its branching fraction,} 
as well as the mass, $M_{K^{*0}}$, and width, $\Gamma_{K^{*0}}$, 
of the $K^{*0}(892)$ resonance\footnote{
We hereafter simply denote the $K^{*0}(892)$ meson as $K^{*0}$.
} are determined.  
%

\vspace*{2 mm}
The measurement was performed 
{with the Belle detector} at the KEKB asymmetric-energy 
$e^+ e^-$ (3.5 on 8 GeV) collider \cite{acce}, using the reaction  
$e^+ e^- \to \tau^+ \tau^-$. 
In this analysis we use a 544.9 fb$^{-1}$ data sample recorded on and near the $\Upsilon$(4S)
resonance, 
corresponding to $5.0\times 10^8$ produced $\tau^{+}\tau^{-}$ pairs. 

The Belle detector is a large-solid-angle magnetic spectrometer that consists of 
a silicon vertex detector (SVD), a 50-layer central drift chamber (CDC), 
an array of aerogel threshold Cherenkov counters (ACC), a barrel-like arrangement 
of time-of-flight scintillation counters (TOF), and an electromagnetic calorimeter (ECL) 
comprised of CsI(Tl) crystals located inside 
a superconducting solenoid coil that 
provides a 1.5 T magnetic field. 
An iron flux-return located outside the coil is instrumented to detect $K_L^0$ mesons 
and to identify muons (KLM). 
The detector is described in detail elsewhere \cite{detec}. 

Particle identification (PID) plays an important 
{r\^ole} in this experiment and 
is based on a likelihood ratio ${\cal P}_x \equiv L_x/\sum_y L_y$ 
for a charged particle $x = \mu$, $e$, $K$, or $\pi$, where the sum runs over the relevant particles. 
$L_x$ is a likelihood based on the energy deposit and shower shape in the ECL, 
the momentum and $dE/dX$ measured by the CDC, the particle range in the KLM, the light 
yield in the ACC, and particle's time-of-flight from the TOF. 
%
\section{Event Selection}

We look for $\tau^-\to K^{*0}(892) K^-\nu_{\tau}$ candidates with the following signature: 
\begin{eqnarray}
\tau^-_{\rm signal} &\to& K^{*0}(892) + K^- + {\rm (missing)} \nonumber \\ 
&~& ~ \hookrightarrow K^+ \pi^- \nonumber\\
\tau^+_{\rm tag} &\to& (\mu/ e)^+ + n (\leq 1)\gamma + {\rm (missing)}. \nonumber
\end{eqnarray}
An event should contain 4 charged tracks with zero net-charge. 
The following basic criteria are imposed; the  
total energy of the tracks and photons in the center-of-mass (CM) frame 
should be less than 11 GeV; 
the missing momentum should be greater than 0.1 GeV/c and lie within the detector acceptance 
$-0.866 < \cos\theta < 0.956$. 
The event is subdivided into two hemispheres according to the thrust axis in the CM frame,  
the signal and tag sides. 
The signal side contains three prongs and no additional photons 
while the one-prong side may include one extra photon.

On the tag side, ${\cal P}_{\mu/e} > 0.1$ is required for 
the track, while on the signal side 
the requirements ${\cal P}_{K} > 0.8$ (${\cal P}_{\pi} > 0.8$)
with ${\cal P}_{e} < 0.9$ are imposed 
for both kaons with opposite charges 
(for the $K^{*0}$ daughter pion) within $\cos\theta > -0.6$.
Here $\theta$ is the polar angle with respect
 to the $z$ axis, which is anti-parallel to the $e^+$ beam direction.
The three signal candidate tracks must have charge assignments that satisfy strangeness conservation i.e. 
$\tau^{\mp}\to K^{*0} (K^{\pm}\pi^{\mp}) K^{\mp}$. 

\begin{figure}[b]
\centerline{
\resizebox{0.46\textwidth}{0.38\textwidth}{%
\includegraphics{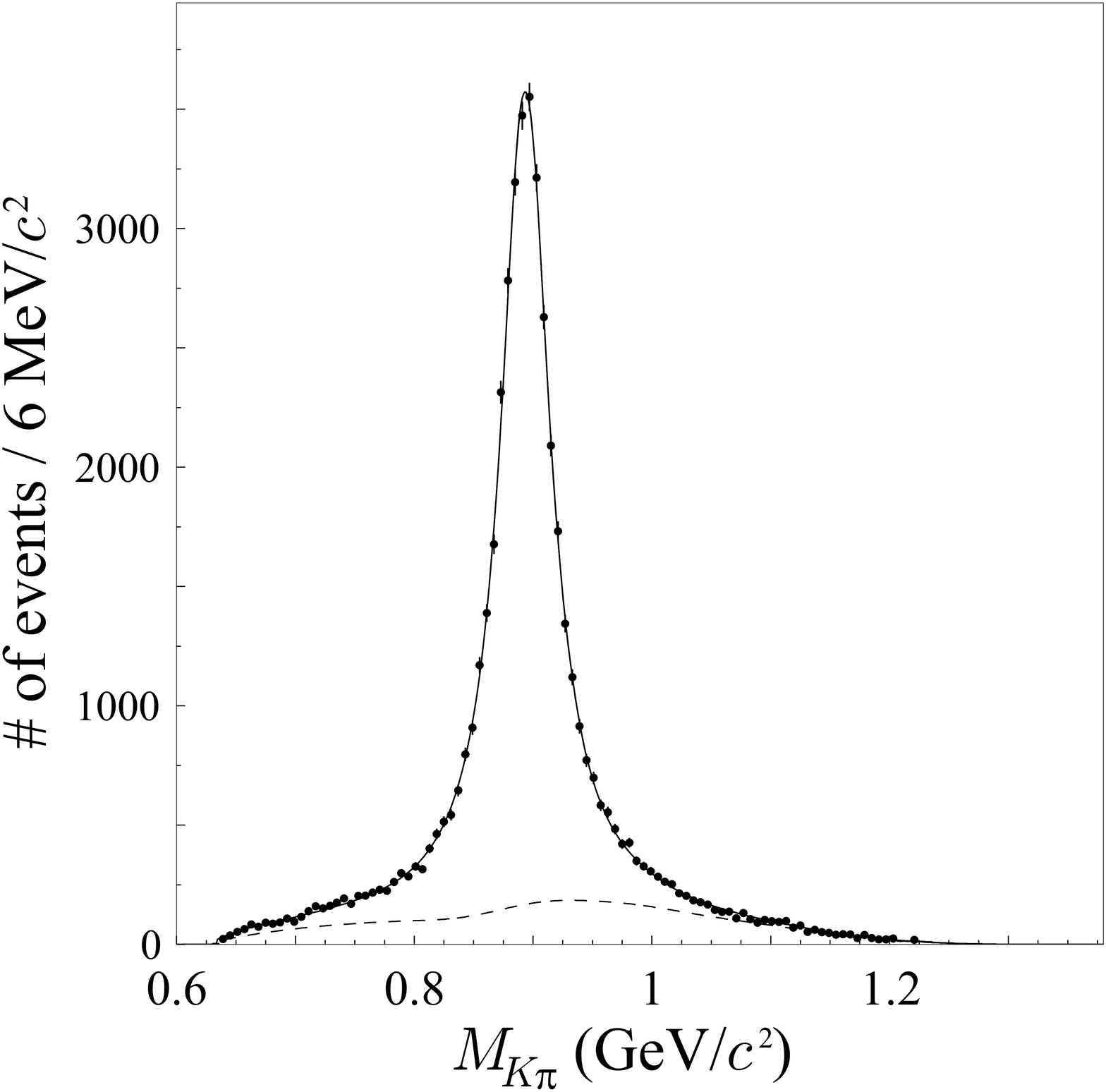}}}
\caption{\small $K^+\pi^-$ mass distribution after the selection. 
Data are indicated by points with error bars. 
The solid curve is the best fit result and the dashed curve is the incoherent BG 
$N^{\rm incoh}(M_{K\pi})$, see the text for details. 
}
\label{F-1}
\end{figure}

In order to reduce $q\overline{q}$ {background}, 
the momenta of the $K^{*0}$ and $K^{*0} K^-$ system are required 
to satisfy the requirements
$p^{\rm CM}_{K^{*0}} > 1.5$ GeV/c and 
$p^{\rm CM}_{K^{*0}K} > 3.5$ GeV/c, respectively; 
the opening angles between the $K^{*0}$ and $K^-$ 
and between the thrust axis and missing 
momentum {should satisfy} the requirements  
$\cos\theta^{\rm CM}_{K^{*0}-K} > 0.92$ and 
$\cos\theta^{\rm CM}_{\rm thrust-miss} < -0.6$, respectively; 
the invariant masses of the particles on both the tag 
and signal sides should be in the ranges  
$M_{\rm tag} < 1.8$ GeV/c$^2$ $(\simeq m_{\tau})$ 
and $M_{\rm signal} < 1.8$ GeV/c$^2$. 

After applying all the selections, 
$5.1\times 10^4$ {events} remain, and  
yield the $K^+\pi^-$ mass ($M_{K\pi}$) distribution,  
shown in Fig.~\ref{F-1}.  
A clear $K^{*0}$ peak is seen. 

The detection efficiency ($\epsilon$) for $\tau^-\to K^{*0}K^-\nu_\tau$ is evaluated using 
the KKMC/PYTHIA Monte Carlo (MC) program \cite{KKMC}, where the V-A interaction is 
assumed for  the weak vertices  
{and all the final state particles} are produced
according to pure phase space. 
For $K^{*0}\to K^+\pi^-$, a spin-dependent  Breit-Wigner (BW) function is used. 
The resulting efficiency, $\epsilon$, is 2.15\% which includes 
the branching fraction 
${\cal B}(K^{*0}\to K^+\pi^-) = 2/3$ 
as well as the tagging efficiency. 

\section{Backgrounds}

The dominant 
background (BG) is from other generic decay modes of $\tau$ lepton pairs. 
The largest component, 
composing $\sim$80\% of the BG,
arises from $\tau^-\to \pi^+\pi^- K^- n\pi^0\nu_\tau$ 
and $\tau^-\to \pi^+\pi^-\pi^- n\pi^0\nu_\tau$ with $n=$0 and 1, through 
$\pi^+\to K^+$ misidentification, 
and an undetected $\pi^{0}$. 
We prepare such enriched BG samples from data by replacing the $K^{+}$ selection criteria 
${\cal P}_K>0.8$ by ${\cal P}_K<0.2$ $(K^+\to \pi^+)$, 
and estimate the contamination  
in the $K^{*0}K^-\nu_\tau$ samples, 
taking into account the PID fake rate.
This estimate has a 1.1\% 
systematic uncertainty, 
mostly due to errors in the PID fake rate.

Another $\sim$20\% of BG is attributed to $\tau^-\to \phi K^-\nu_\tau$, 
$K^+ \pi^- K^- \pi^0 \nu_\tau ~({\rm excluding}\ K^{*0})$, 
$\phi\pi^-\nu_\tau$, and $K^+ \pi^- K^- \nu_\tau ~({\rm ex.}~K^{*0})$,
and is estimated by MC simulation.
The first and second background components arise from  mis-PID $(K^-\to \pi^-)$ and 
missing $\pi^0$ detection, respectively, at rates of  $(1.4\pm 0.2)\%$ and 
$(2.1\pm 0.7) \%$. 
Their $M_{K\pi}$ spectra are evaluated by MC, based on their branching fractions, 
reported in 
{Refs.~\cite{phiK} and \cite{PDG}}, 
with uncertainties of $0.07\%$ and $0.6\%$ of the $K^{*0}$ yields.

{The third background component has the same final state, 
$K^+\pi^-K^-\nu_\tau$, as that of $ K^{*0} K^-\nu_\tau$,
and its $M_{K\pi}$ spectrum ($N^{\phi\pi^-\nu}(M_{K\pi})$)
is obtained in \cite{phiPi}.}
The fourth BG is the non-resonant (NR) component and
{can interfere} with the signal. 
Details are discussed in the next section.

%

\begin{figure}[h]
\centerline{
\resizebox{0.45\textwidth}{0.35\textwidth}{%
\includegraphics{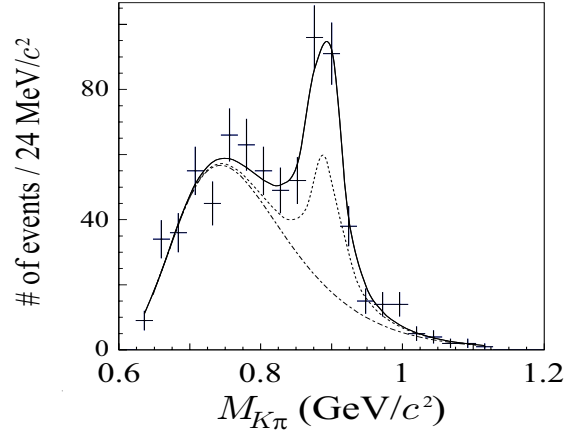}}}
\caption{\small $K^+\pi^-$ mass distribution 
for $K^{*0} K^-\pi^0\nu_\tau$ candidates. 
Data are plotted by crosses, and the best-fit result is indicated 
by {the} solid curve, 
{the} $K^{*0} K^-\nu_\tau$ contamination and 
the {non-$K^{*0}$ BG} 
{are shown by the dotted curve}
and dash-dotted one, respectively. 
{See the text for details.}
}
\label{F-2}
\end{figure}

On the other hand, $K^{*0}$-resonant BG comes from as yet unmeasured 
$\tau^-\to K^{*0} K^- \pi^0\nu_\tau$ decay and $q\overline{q}$ processes. 
In order to measure ${\cal B}(\tau^-\to K^{*0}K^-\pi^0\nu_\tau)$, 
we apply the same criteria {as}
for the $K^{*0} K^-\nu_\tau$ selection, 
but with additional requirements of $n_{\gamma}=2$ 
and effective mass ($M_{\gamma\gamma}$) in the range,
$0.1178$ GeV/c$^2$ $< M_{\gamma\gamma} < 0.1502$ GeV/c$^2$.
Figure \ref{F-2} shows the resulting  $M_{K\pi}$ distribution, 
where the $K^{*0}$ peak includes  two contributions: 
one is the $K^{*0} K^-\pi^0\nu_\tau$ signal, 
and {the other comes from} $K^{*0} K^-\nu_\tau$ events 
with a $\pi^0$ formed by spurious photons, caused 
by a splitoff of  a hadronic shower 
in the ECL calorimeter. 
The latter contribution is calculated by MC, using ${\cal B}(\tau^-\to K^{*0} K^-\nu_\tau)$ 
from the PDG \cite{PDG} and our result. 
The {non-$K^{*0}$ BG}
in Fig.~\ref{F-2} mostly arises from $\tau^+\tau^-$ pair processes, 
while {the} $q\overline{q}$ contribution is negligibly small. 
The $M_{K\pi}$ distribution is fitted with a $K^{*0}$ BW function plus 
a {non-$K^{*0}$ BG} represented by a Landau function, 
with four free parameters: 
the total number of $K^{*0} K^-\pi^0\nu_\tau$ events, and three  
parameters of the Landau function. 
The best fit yields the number of {signal events}
as $N_{K^{*0} K^-\pi^0\nu}= 129.2\pm 25.1$, 
with a detection efficiency $\epsilon = 0.54\%$,
fixing the $K^{*0}K^-\nu_\tau$ peaking BG to  
$N_{K^{*0} K^-\nu}= 113.7\pm 6.8$ events. 
As a result, we obtain the first measurement of the branching fraction,  
\begin{eqnarray}
\label{eq-1}
{\cal B}(\tau^-\to K^{*0} K^-\pi^0\nu_\tau) = \hspace*{3 cm} \nonumber \\
(2.39\pm 0.46\pm 0.26)\times 10^{-5}, \ \
\end{eqnarray}
%
with a systematic uncertainty of 11.0\%.
Table~\ref{T-1} lists 
the {sources} of individual uncertainties. 
The largest contribution, 9.0\%, arises from an uncertainty 
{on the peaking BG estimate, especially, 
on $\tau^-\to K^{*0} K^-\nu_\tau$. }
This $K^{*0} K^-\pi^0\nu_\tau$ contamination in the $K^{*0} K^-\nu_\tau$ signal in the 
$M_{K\pi}$ distribution is only 0.4\%. 
%

The other possible $K^{*0}$-resonant BG, the $q\overline{q}$ contribution, mostly from $uds$, 
is examined in a comparison between data and MC, by applying $q\overline{q}$ enriching 
selection criteria. 
The estimated contribution in the $K^{*0}$ mass region is $46 \pm 17$ events:  
It gives an uncertainty of only 0.14\% to the estimation 
of the signal events for the $K^{*0}K^-\nu_\tau$ analysis. 

\section{\boldmath ${\cal B}(\tau^-\to K^{*0} K^-\nu_\tau)$}

\begin{table*}[t]
\begin{center}
\caption{Results of fits over the mass region of $0.645-1.147$ (GeV/c$^{2}$) . 
'standard' means the fit with five free parameters ($\alpha$, $\beta$, 
$M_{K^{*0}}$, $\Gamma_{K^{*0}}$ and $\phi$) with 
{$r=3.53$,} and with $r=0$. 
The interference between $A_{BW}$ and $A_{NR}$ is 
{excluded} in the 'no interference' case, 
and then $\phi$ is removed from the fit. 
In the rightmost column, 
$A_{NR}$ is omitted in the fit. 
To obtain branching fractions, we use detection efficiencies, as
$\epsilon_{K^* K}= (2.15\pm 0.01) \%$ and 
$\epsilon_{KK\pi} = (2.60\pm 0.01) \%$, determined from MC simulations 
based on pure hadronic phase space with a $V-A$ weak interaction. 
In ${\cal B}(\tau^-\to K^{*0} K^-\nu_\tau)$, ${\cal B}(K^{*0}\to K^+\pi^-)=2/3$ is included. }
\label{T-2}
{
\begin{tabular}{|c|rr|r|r|} \hline
 & \multicolumn{2}{c|}{standard} & \multicolumn{1}{c|}{no interference} & 
 \multicolumn{1}{c|}{no $A_{NR}$}  \\ \hline 
 & $r=3.53~~~$ & $r=0.0~~~$ & $r=3.53~~~$ & $r=3.53~~~$ \\ \hline
$M_{K^{*0}}$ (MeV/c$^2$) & $895.25 \pm 0.27$ & $896.58\pm 0.28$ 
& $896.72\pm 0.19$ & $896.85\pm 0.19$  \\
$\Gamma_{K^{*0}}$ (MeV/c$^2$) & $47.70\pm 0.49$ & $48.19\pm 0.51$ 
& $47.45\pm 0.48$ & $49.21\pm 0.46$  \\
$\phi$ $(^\circ)$ & $63.46\pm 2.05$ & $37.98\pm 2.90$ & $-~~~~~~$ & $-~~~~~~$  \\
${\cal B}(K^{*0} K^-\nu)$~~$(\times 10^{-3})$ & $1.56\pm 0.02$ & $1.55\pm 0.02$ 
& $1.80 \pm 0.01$ & $1.84 \pm 0.01$  \\
${\cal B}(K^+\pi^- K^-\nu)$~~$(\times 10^{-5})$ & $5.76\pm 0.59$ & $4.62\pm 0.53$ 
& $4.82\pm 0.54$ & $-~~~~~~$  \\ \hline
$\chi^2/{\rm ndf}$ & $83.29/80~~~$  & $84.64/80~~~$ & 
    $146.3/81~~~$ & $226.2/82~~~$ \\ \hline
\end{tabular}
}
\end{center}
\end{table*}

The $M_{K\pi}$ distribution in Fig.~\ref{F-1} is fitted with 
the following formula: 
\begin{eqnarray}
N(M_{K\pi}) = \Big{|} 
\alpha A_{BW}(M_{K\pi})
+ \beta A_{NR}(M_{K\pi}) e^{i\phi} 
\Big{|}^2 
\nonumber \\
+ N^{\rm \phi \pi^-\nu}(M_{K\pi}) 
+ N^{\rm incoh}(M_{K\pi}) \ \
\end{eqnarray}
where $A_{BW}(M_{K\pi})$ is the Breit-Wigner (BW) function for 
the decay of a $J^P=1^- \to 0^- 0^-$ state, expressed as 
\begin{eqnarray}
A_{BW}(M_{K\pi}) = \frac{M_0 \Gamma}
{(M^2_0-M^2_{K\pi}) - i M_0 \Gamma} , ~~~~~~\nonumber\\
\Gamma(M_{K\pi}) = \left(\frac{q}{q_0}\right)^3 
\left(\frac{M_0}{M_{K\pi}}\right)
\frac{D_L(q_0 r)}{D_L(qr)} \Gamma_0 . \ \ 
\end{eqnarray}
$M_0$ and $\Gamma_0$ are the $M_{K^{*0}}$ and $\Gamma_{K^{*0}}$, respectively. 
$q$ is the momentum in the $K^+\pi^-$ center of mass, and 
$D_L(q r) = 1/(1+r^2 q^2)$ is the barrier factor with the so-called damping factor $r$. 
Since the damping factors for $K^{*0}(892)\to K^+\pi^-$ decay 
were obtained by LASS \cite{LASS} in the $K^- p\to K\pi n$ reaction,
$r = 3.4\pm 0.6\pm 0.3$ (GeV/c)$^{-1}$, 
and by FOCUS \cite{FOCUS} in $D^+\to K^-\pi^+\mu^+\nu$ decay, 
$r= 3.96 \pm 0.54 {}^{+1.31}_{-0.90}$ (GeV/c)$^{-1}$, 
we take the average of these two values, 
$r= 3.53\pm 0.59$ (GeV/c)$^{-1}$ in the fits. 
The term $A_{\rm NR}(M_{K\pi})$ is the non-resonant $\tau^-\to K^+\pi^- K^-\nu_\tau$ 
scalar amplitude, and can be expressed as 
\begin{eqnarray}
A_{NR} = \left(\frac{M_{K\pi}}{q}\right) \sin\delta_{LASS} e^{i\delta_{LASS}}
\end{eqnarray}
where the phase-shift of $\delta_{LASS}$ is parametrized as 
\begin{eqnarray}
\cot\delta_{LASS} = \frac{1}{aq} + \frac{bq}{2}
\end{eqnarray}
following the LASS \cite {LASS} and FOCUS \cite{FOCUS} analyses with 
$a=4.03\pm 1.72\pm 0.06$ (GeV)$^{-1}$ and $b=1.29\pm 0.63\pm 0.67$ (GeV)$^{-1}$, 
obtained by LASS. 
The phase $\delta_{LASS}$ has a weak $q$-dependence and is $\simeq 45^\circ$ in our  
momentum range. 
{Since the final hadronic system contains 
an additional $K^-$, final state interactions are possible between
$K^{*0}$ and non-resonant $K^+\pi^-$ systems and $K^-$}, and
a relative phase between the $A_{BW}$ and $A_{NR}$ 
amplitudes is introduced in the formula. 
Here, $N^{\rm \phi \pi^-\nu}(M_{K\pi})$ is the contribution from
$\phi \pi^-\nu_\tau$, 
and $N^{\rm incoh}(M_{K\pi})$ is 
the sum of the incoherent continuum and peaking BG's discussed 
in the previous section.  
{These correspond to 
$795\pm65$ and $10,073 \pm 994$ events, respectively.}
The free parameters in the $\chi^2$-fits are $\alpha$, $\beta$, $M_{K^{*0}}$, 
$\Gamma_{K^{*0}}$, and $\phi$.

The mass dependence of the resolution 
{is taken into account} in the fit. 
The resolution function can be approximated as a sum of three Gaussian functions 
with their relative fractions and 
{standard deviations $\sigma$'s:}  
about 52\% with $1.7$ MeV/c$^2$, 38\% with $3.4$ MeV/c$^2$ and 10\% 
with $9.2$ MeV/c$^2$ 
at $M_{K\pi}\simeq 890$ MeV/c$^2$; 
$\sigma$'s vary 
by $3.7\times 10^{-3}$, $5.6\times 10^{-3}$ and 
$3.3 \times 10^{-3}$ 
{per MeV/c${}^2$}
with $M_{K\pi}$, respectively; 
the third Gaussian is shifted to the higher mass side by $\simeq 0.13$ MeV/c$^2$ 
relative to the others.

%

\vspace*{2 mm}

\begin{figure}[h]
\centerline{
\resizebox{0.45\textwidth}{0.35\textwidth}{%
\includegraphics{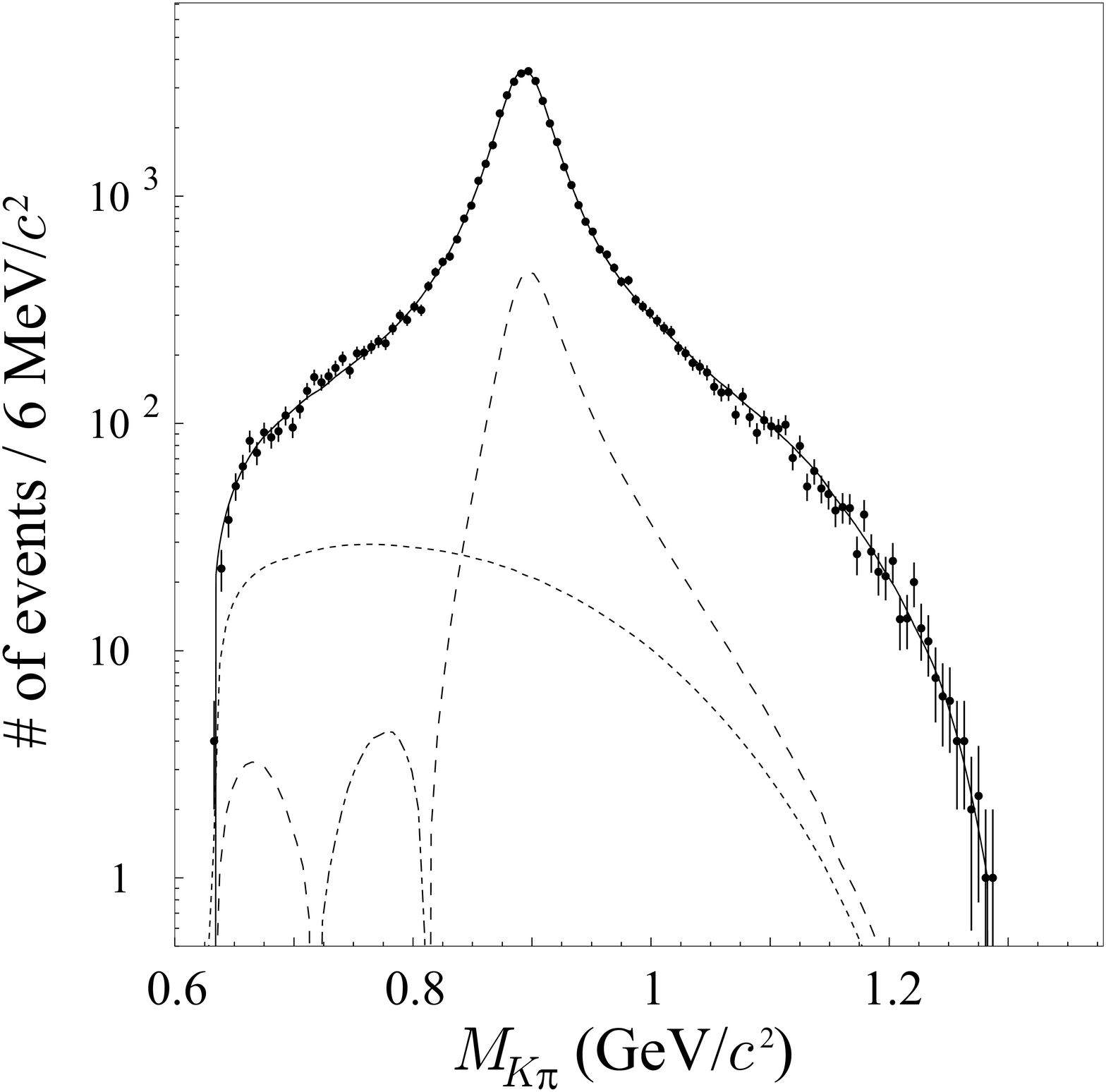}}}
\caption{Result of the best-fit, plotted 
{with} a logarithmic scale. 
The spectra of all terms, 
the interference term and the non-resonant term 
are plotted by the solid, dashed and dotted curves, respectively. 
The negative interference term is reversed and plotted by 
the dash-dotted curve. 
See Fig.~\ref{F-1} for the same plot, 
but {with} a linear scale.  
}
\label{F-3}
\end{figure}

The $\chi^2$-fits are performed in the mass range 
$M_{K\pi}=0.645-1.147$ (GeV/c$^2$). 
The results of fits are listed in Table~\ref{T-2}, and 
the best fit is shown in Figs. \ref{F-1} and \ref{F-3}. 
The branching fractions obtained are 
${\cal B}(\tau^-\to K^{*0} K^-\nu_\tau)=(1.56\pm 0.02)\times 10^{-3}$ 
and ${\cal B}(\tau^-\to K^+ \pi^- K^-\nu_\tau)_{\rm non-resonant}
=(5.76\pm 0.59)\times 10^{-5}$. 
The non-resonant $K^+\pi^- K^-\nu_\tau$
contribution is 
$(5.58\pm 0.57) \%$, which agrees well with the FOCUS result of 
$(5.30\pm 0.74^{+0.99}_{-0.96}) \%$.

{
Since $\phi + \delta_{LASS} \simeq 65^\circ + 45^\circ \simeq 110^\circ$, }
the interference term $2 \alpha \beta A_{BW} A^*_{NR}$ has a  $M_{K\pi}$ distribution 
similar to a BW shape, as seen in Fig.~\ref{F-3}. 
The role of the damping factor in the $M_{K\pi}$ distribution 
in this reaction resembles that of $\phi$ 
in the interference term on the distribution. 
This can be seen in the $\phi=0^\circ$ case,
where the parameters resulting from the fit, 
except for $\phi$, do not change much 
compared to the $r=3.53$ case,
while $\phi$ changes from $(37.98\pm 2.90)^\circ$ to 
{$(63.46\pm 2.05)^\circ$}
with a $\chi^2$ difference of 
{$\Delta \chi^2=1.35$.}


When the interference contribution is not taken into account, 
${\cal B}(\tau^-\to K^{*0} K^-\nu_\tau)$ becomes 15\% larger 
(see the fourth column in Table~\ref{T-2}); 
when the NR contribution is totally ignored, 
${\cal B}(\tau^-\to K^{*0} K^-\nu_\tau)$ 
becomes 2.2\% larger than 
{in the case} where
the interference contribution is not taken into account
(see the rightmost column in Table~\ref{T-2}).

The systematic uncertainties are listed in Table~\ref{T-1}.
The total uncertainty is  5.5\%, 
where the largest errors are attributed to the track finding efficiency, 
and PID's on the lepton and {kaon}, 
as 3.3\%, 2.9\% and 2.6\%, respectively. 
The $r$ parameter is varied by $\pm 1\sigma$ 
to evaluate its effect on the branching fractions: 
the resulting change is $\pm 0.3\%$
for ${\cal B}(\tau^-\to K^{*0} K^-\nu_\tau)$ and 
$\pm1.6\%$
for ${\cal B}(\tau^-\to K^+\pi^- K^-\nu_\tau)$. 
These systematic uncertainties are much smaller than 
{the statistical error of the fit}. 

\begin{table}[b]
\begin{center}
\caption{\small Systematic errors for ${\cal B}(\tau^-\to K^{*0} K^-\nu_\tau)$ (left column) 
and ${\cal B}(\tau^-\to K^{*0} K^- \pi^0 \nu_\tau)$ (right column) in \%. }
\label{T-1}
\begin{tabular}{|l|cc|} 
\hline 
 & $K^{*0} K^-\nu$ & $K^{*0} K^-\pi^0\nu$ \\
\hline 
Luminosity & 1.4 & 1.4 \\
$\sigma(e^+ e^-\to \tau\tau)$ & 0.3 & 0.3 \\
Tracking efficiency & 3.3 & 3.3 \\
Trigger efficiency & 0.7 & 0.1 \\
Lepton-ID & 2.9 & 2.9 \\
Kaon-ID/fake & 2.6 & 3.8 \\
MC statistics & 0.3 & 0.5 \\
$\pi^0$ efficiency & -- & 1.7 \\
BG estimate & 1.3 & 9.0 \\
\hline
Total & 5.5 & 11.0 \\
\hline
\end{tabular}
\end{center}
\end{table}

Adding all systematic errors in quadrature, the branching fraction obtained is 
\begin{eqnarray}
\label{eq-2}
{\cal B}(\tau^-\to K^{*0} K^-\nu_\tau) = \hspace*{3 cm} \nonumber\\
(1.56 \pm 0.02 \pm 0.09) \times 10^{-3}.\ \ 
\end{eqnarray}

The branching fraction of non-resonant $\tau^-\to K^+\pi^-K^-\nu_\tau$
is also obtained, 
\begin{eqnarray}
\label{eq-3}
{\cal B}(\tau^-\to K^+\pi^-K^-\nu_\tau)_{\rm non-resonant} 
= \hspace*{1 cm} \nonumber \\
(5.76\pm 0.59\pm 2.04)\times 10^{-5}.\ \
\end{eqnarray}
This is the first measurement of non-resonant $\tau^-\to K^+\pi^-K^-\nu_\tau$ 
decay, where the large systematic error mostly comes from 
{the uncertainty} in estimating 
the $\pi^+\pi^-\pi^-\nu_\tau$ BG contamination through mis-PID.

\section{\boldmath $M_{K^{*0}}$ and $\Gamma_{K^{*0}}$}

The absolute mass scale for the Belle detector is confirmed 
by reconstructing {the} $K_S$ 
and $\phi$ masses through their $\pi^-\pi^+$ and $K^- K^+$ decay modes, 
respectively, and comparing them to their world average values.
Such studies provide mass differences from the PDG values \cite{PDG} of  
$-0.045\pm 0.027$ MeV/c$^2$ and $-0.14\pm 0.20$ MeV/c$^2$ for $M_{K_S}$ and 
$M_{\phi}$, respectively, or from the most precise measurements 
\cite{NA48} and \cite{CMD2}, 
$-0.022\pm 0.035$ MeV/c$^2$ and $-0.12\pm 0.20$ MeV/c$^2$. 
Therefore, no correction is applied to the absolute mass scale.

On the other hand, MC simulation results in shifts of 
$\Delta M_{K^{*0}} = 0.15\pm 0.04$ MeV/c$^2$ 
and $\Delta \Gamma_{K^{*0}} = 0.47 \pm 0.22$ MeV/c$^2$, 
due to the effects of detection efficiency and event selections, 
and therefore we correct the raw $M_{K^{*0}}$ and $\Gamma_{K^{*0}}$ 
{values}
by these amounts, respectively.

\begin{table*}[thb]
\begin{center}
\caption{Comparison of our results with other experiments}
\label{table-2}
\begin{tabular}{|l|l|l|} \hline\hline
    & $M_{K^{*0}(892)}$ (MeV/c$^2$)& $\Gamma_{K^{*0}(892)}$ (MeV/c$^2$)
 \\ \hline
Belle     &  
$895.10\pm 0.27\pm 0.31$ &
$47.23\pm 0.49 \pm 0.79$   \\
PDG \cite{PDG}     &              $896.00\pm 0.25$              & $50.3\pm 0.6$   \\
FOCUS \cite{FOCUS} & $895.41\pm 0.32\pm ^{0.35}_{0.43}$ 
         & $47.79\pm 0.86\pm ^{1.32}_{1.06}$  \\
LASS  \cite{LASS}  & $895.9\pm 0.5\pm 0.2$  & $50.8\pm 0.8 \pm 0.9$   \\
\hline
    & $M_{K^{*\pm}(892)}$ (MeV/c$^2$)& $\Gamma_{K^{*\pm}(892)}$ (MeV/$c^2$) \\ \hline
Belle \cite{Epif} & $895.47\pm 0.20\pm 0.74$ & $46.2\pm0.6\pm1.2$\\
PDG \cite{PDG} &  $891.66\pm 0.26$ & $50.8\pm 0.9$ \\
\hline\hline
\end{tabular}
\end{center}
\end{table*}

To evaluate uncertainties in the BG estimates on $M_{K^{*0}}$ and $\Gamma_{K^{*0}}$, 
we vary the mis-PID probabilities by $\pm 1 \sigma$ in the fit. 
The resulting variation is 
{$\delta M_{K^{*0}}=\pm 0.23$MeV/c$^2$}
and {$\delta \Gamma_{K^{*0}}=\pm0.75$ MeV/c$^2$.}
The uncertainty in $r$ is included by varying it by $\pm 1\sigma$ in the fit, 
which gives changes of {$M_{K^{*0}}=\pm 0.22$ MeV/c$^2$} and 
{$\Gamma_{K^{*0}}=\pm 0.09$ MeV/c$^2$. }

Including the above errors in quadrature, the mass and width are
\begin{eqnarray}
\label{eq-4}
M_{K^{*0}} &=& 895.10 \pm 0.27 \pm 0.31  \ {\rm (MeV/c^2)}, \ \ \\
\label{eq-5}
\Gamma_{K^{*0}} &=& 47.23 \pm 0.49 \pm 0.79 \ {\rm (MeV/c^2)},\ \ 
\end{eqnarray}
where the first and second errors are statistical 
and systematic ones, respectively. 

\section{Conclusions}

With the use of {high-statistics} data samples of 
{544.9} fb$^{-1}$, 
{collected in the}
Belle experiment, 
we have measured ${\cal B}(\tau^-\to K^{*0} K^-\nu_\tau)$ with 
the highest precision so far attained, 
and obtained the first measurements of 
${\cal B}(\tau^-\to K^{*0} K^-\pi^0\nu_\tau)$ and 
${\cal B}(\tau^-\to K^+ \pi^-K^-\nu_\tau)_{\rm non-resonant}$, 
respectively. 

{The ${K^{*0}}$ mass and width} are also measured, 
respectively, with better or 
{same accuracy} than those of 
the PDG world {average} values, 
or the most precise previous experiments, as listed in 
Table~\ref{table-2}.

\begin{figure}[bht]
\centerline{
\resizebox{0.4\textwidth}{0.4\textwidth}{%
\includegraphics{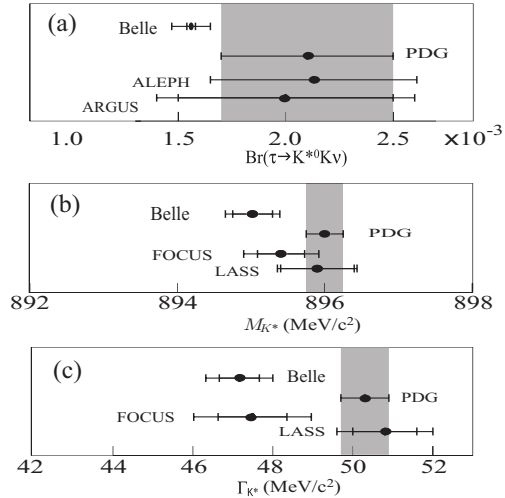}}}
\caption{
{Comparison of various measurements:}
(a) ${\cal B}(\tau^-\to K^{*0} K^-\nu_\tau)$, 
(b) $M_{K^{*0}}$ and (c) $\Gamma_{K^{*0}}$.
In (a), the PDG value is obtained from ARGUS~\cite{ARGUS} 
and ALEPH~\cite{ALEPH}. 
In (b) and (c), FOCUS and LASS results are included in the PDG value.}
\label{F-4}
\end{figure}

Figure~\ref{F-4}
compares various measurements:
our results differ from the PDG world average values, 
but are quite consistent with the results of the FOCUS experiment 
for $M_{K^{*0}}$ and $\Gamma_{K^{*0}}$. 
The difference between our measurement 
of ${\cal B}(\tau^-\to K^{*0} K^-\nu_\tau)$ and 
other results may 
be due to the inclusion of the interference effect 
between the decays 
that have the same final state $K^{+}\pi^{-}K^{-}\nu_{\tau}$.

{\ }\\
\noindent
{\bf Acknowledgements}\\

We thank the KEKB group for the excellent operation of the
accelerator, the KEK cryogenics group for the efficient
operation of the solenoid, and the KEK computer group and
the National Institute of Informatics for valuable computing
and SINET3 network support. We acknowledge support from
the Ministry of Education, Culture, Sports, Science, and
Technology of Japan and the Japan Society for the Promotion
of Science; the Australian Research Council and the
Australian Department of Education, Science and Training;
the National Natural Science Foundation of China under
contract No.~10575109 and 10775142; the Department of
Science and Technology of India; 
the BK21 program of the Ministry of Education of Korea, 
the CHEP SRC program and Basic Research program 
(grant No.~R01-2005-000-10089-0) of the Korea Science and
Engineering Foundation, and the Pure Basic Research Group 
program of the Korea Research Foundation; 
the Polish State Committee for Scientific Research; 
the Ministry of Education and Science of the Russian
Federation and the Russian Federal Agency for Atomic Energy;
the Slovenian Research Agency;  the Swiss
National Science Foundation; the National Science Council
and the Ministry of Education of Taiwan; and the U.S.\
Department of Energy.

\end{document}